\documentclass[aps,twocolumn,nofootinbib,preprintnumbers,superscriptaddress]{revtex4-1}
\pdfoutput=1
\usepackage[usenames,dvipsnames,table]{xcolor}
\usepackage{soul}
\usepackage{graphicx,amsmath,amssymb,amsthm,multirow,array,bm}
\usepackage[mathscr]{eucal}
\usepackage{amsfonts}
\usepackage{hyperref}
\usepackage{tikz}
\usepackage{bbm}
\usepackage{ulem}
\usepackage{lipsum}
\usepackage{datetime}

\usepackage[utf8]{inputenc}

\def\red#1{\textcolor{black}{#1}}

\def\nuen#1{\textcolor{black}{#1}}

\hypersetup{
    pdfstartview={FitH},    
    pdftitle={},    
    pdfauthor={Diptarka Das, Shouvik Datta, Sridip Pal},     
    colorlinks=true,       
    linkcolor=blue,          
    citecolor=red!50!black,        
    filecolor=magenta,      
    urlcolor=blue           
}

\newcommand\underrel[2]{\mathrel{\mathop{#2}\limits_{#1}}}
\definecolor{rust}{rgb}{0.8,0.2,0.2}

\def\vev#1{\langle\, #1 \, \rangle}
\def\ket#1{\mid \! #1\rangle}

\def\nn{\nonumber}

\def\l1{{\text{1-loop}}}

\def\bz{{\bar{z}}}

\def\n1{\Bigg|_{n=1}}

\def\n{{(n)}}

\def\O{\mathcal{O}}

\def\ket#1{|#1\rangle}

\def\vev#1{\langle{#1}\rangle}

\def\cS{\mathcal{S}}

\def\tq{\tilde{q}}
\def\c{\mathcal{C}}

\def\cS{\mathcal{S}}

\def\cF{{\mathcal{F}}}
\def\cV{{\mathcal{V}}}
\def\ie{{\it i.e.~}}

\begin{document}

\title
{ 
 Modular crossings, OPE coefficients and black holes  
}
 \preprint{}




\author{Diptarka Das}
\email{ddas@aei.mpg.de}
\affiliation{Max Planck Institut f\"ur Gravitationsphysik, Potsdam-Golm, D-14476, Germany.}

\author{Shouvik Datta}
\email{shouvik@itp.phys.ethz.ch}
\affiliation{Institut f\"ur Theoretische Physik,  ETH Zurich, CH-8093 Z\"urich, Switzerland. }

\author{Sridip Pal}
\email{srpal@ucsd.edu}
\affiliation{Department of Physics, University of California at San Diego, La Jolla, CA 92093, USA.}

\begin{abstract}

In (1+1)-$d$ CFTs, the 4-point function on the plane can be mapped to the pillow geometry and thereby crossing symmetry gets translated into a modular property. We use these modular features to derive a universal asymptotic formula for OPE coefficients in which one of the operators is averaged over heavy primaries.  The coarse-grained heavy channel then reproduces features of the  gravitational 2$\to$2 $\cS$-matrix which has black holes as their intermediate states.

\end{abstract}


\maketitle

\section{Introduction}
\label{sec:intro}

 Conformal field theory (CFT) pervades several areas of theoretical physics today.
    ~Amongst its varied  uses, it appears in systems nearing phase transitions \cite{Cardy:1989da}, describes quantum impurities \cite{affleck1995conformal}, portrays  string worldsheets and serves as the holographic equivalent to quantum gravity in Anti-de-Sitter space \cite{Maldacena:1997re}. A CFT is uniquely characterized by its spectrum of primaries and OPE coefficients. In two dimensions, there is an infinite dimensional enhancement of conformal symmetry. Furthermore, when placed on a torus,   modular invariance of CFTs leads to additional constraints \cite{Hellerman:2009bu,Keller:2012mr,Friedan:2013cba,Qualls:2013eha,Collier:2016cls}. These constraints have led to universal properties of the  spectrum \cite{Cardy:1986ie} and more recently heavy-heavy-light OPE coefficients \cite{km,Das:2017vej}. For holographic CFTs, the high-energy asymptotics (combined with a coarse graining of the heavy microstates) reproduce features of black holes in AdS. One of the \red{well-known} examples is that of Cardy's formula \cite{Cardy:1986ie} which gives the Bekenstein-Hawking entropy of the BTZ black hole \cite{Strominger:1997eq,Carlip:2005zn} \red{(which, in turn, retains its validity in an extended energy regime  \cite{Hartman:2014oaa}).}

 The conformal bootstrap programme is aimed at utilizing crossing symmetry of correlators to pin down the CFT data \cite{Ferrara:1973eg}. This has met with great successes over the last few years \cite{Rattazzi:2008pe,ElShowk:2012ht}.
 In a recent development \cite{MSZ}, a novel method has been prescribed to translate crossing symmetry  of CFT$_2$ 4-point functions to a modular property using the structure of the Virasoro blocks \cite{Zamu}.
\red{This feature had also been pointed out earlier in the context of the Ashkin-Teller model  \cite{Zamolodchikov:1985}.} 
 In this work, we \red{utilize this modular property to extract} the mean-squared 
  OPE coefficient, in which one index is averaged over heavy primaries.
  This information is then used to evaluate the contribution of the coarse-grained heavy channel  to the 4-point function of primaries.   \red{For CFTs fulfilling the criteria to admit a gravity dual}, this coarse-grained rendition of heavy microstates holographically corresponds to a black hole. 
   We shall demonstrate that our CFT analysis provides a precise derivation of the holographic 2$\to$2 $\cS$-matrix which has black holes as their intermediate states. It agrees with previous results and expectations in the literature \cite{GiddingsPorto,Giddings:2007qq,FK-S-matrix,RychkovOPE,ArkaniHamed:2007ky,Dvali:2014ila} {i.e.~}this amplitude is entropically suppressed as  $\exp({-S_{\rm BH}}/2)$, where $S_{\rm BH}$ is the Bekenstein-Hawking entropy.
 
 A direct and full-fledged analysis of a scattering process with black hole resonances is a formidable problem in quantum gravity. Amongst its many subtleties, we need to work with a specific UV completion, find proper ways to regulate divergences, tackle the resummation of loop diagrams and, most importantly, be wary about issues regarding unitarity and information loss (associated with the process of creation and evaporation of black holes) \cite{Hawking:1976ra,Almheiri:2012rt,Harlow:2014yka}. However, if holographically mapped, the CFT$_2$ version of the problem is tractable non-perturbatively and stands robust against unitarity concerns \cite{Fitzpatrick:2016ive,Chen:2017yze,Anous:2016kss}. 
 {This CFT analysis may also offer clues for studying the process in AdS$_3$ gravity. }
 Although this is a rather simple setting to address these questions, we hope that it sheds light on analogues of the problem in higher dimensions, as it already captures some of the most important characteristics of the $\cS$-matrix which are expected.

 Apart from these holographic implications, our findings sharpen the notions of OPE convergence within the very  structure of CFTs \cite{Mack:1976pa,RychkovOPE,FK-S-matrix}. As is well known, modular invariance requires an infinite number of primaries for CFTs with Virasoro symmetry and having central charge greater than one \cite{Cardy:1986ie}. It is therefore of pivotal importance to verify the convergence of OPEs and it is reassuring to see that this expectation is 
 indeed true. 
 
%

\section{Modular properties of \\ the 4-point function}
\label{sec:mod}
Consider a  2-dimensional CFT with central charge, $c>1$, which has Virasoro symmetry as its chiral algebra. The basic object, we want to look at, is the four point function of identical scalar primaries of dimension $\Delta_\O$, 
\begin{align}
\cF(z,\bz) = \vev{ \O(0) \O(z,\bz) \O(1)  \O(\infty)} .
\end{align}
Here, $z,\bar{z}$ is the cross-ratio. Crossing symmetry, which is the statement about associativity of operator product expansions, implies
\begin{align}\label{cross1}
\cF(z,\bz) = \cF(1-z,1-\bz). 
\end{align}
Inserting a complete set of states, $\cF(z, \bz )$ can be decomposed in terms of Virasoro conformal blocks $\cV_h(z)$, 
\begin{align}\label{cF}
\cF(z,\bz) = \sum_{h,\bar{h}} \cF_{h,\bar{h}} (z,\bz) \equiv \sum_{h,\bar{h}} f^2_{\O \O \O_{h,\bar{h}}} \cV_h(z) \cV_{\bar{h}}(\bz).
\end{align}
Each term, $ \cF_{h,\bar{h}} (z,\bz)$, in the above sum calculates the `partial wave amplitude' of the exchange channel labelled by the primary of the corresponding Virasoro block. 
The structure of the Virasoro conformal blocks, $\cV_h(z)$, are however quite intricate \cite{Zamu}, and it turns out that an alternate representation of $\cF(z,\bz)$ is more useful. The function $\cF(z,\bz)$ is defined over a Riemann sphere which is marked at the operator locations i.e. at $0, z,1$ and $\infty$. This presentation can be equivalently depicted as a $\mathbb{Z}_2$-quotient of the torus, $\mathbb{P}^1\equiv\mathbb{T}^2/\mathbb{Z}_2$, commonly referred to as the `pillow' geometry, Fig.~1 \cite{MSZ}. The transformation from the sphere to the pillow yields the elliptic representation of the 4-point function. The nome $q$ which appears in this representation is given by $q= e^{i \pi \tau}$, where the modular parameter $\tau$, is related to the cross ratio by the relation $\tau = i K(1-z)/K(z)$ and $K(z)$ is an elliptic integral of the first kind. Note that crucially this implies, that taking $z \rightarrow 1-z$ is equivalent to the S-modular transformation, $\tau \rightarrow -1/\tau$ or $q\rightarrow \tilde{q}$. The pillow geometry makes these modular features manifest.


In the pillow frame, $\mathbb{P}^1$, the operators are located at the four fixed points as indicated in Fig.~1. Taking into account local rescalings from these insertion points as well as the Weyl anomaly associated with the change of conformal frame, we can express the original 4-point function as, 
\begin{align}\label{relate}
{\cal F}(z, \bar z) = \Lambda(z)\Lambda(\bar z) g(q , \bar q).
\end{align}
Here $\Lambda(z) \equiv \vartheta_3(q)^{{c\over 2}-8\Delta_\O}(z(1-z))^{{c\over 24}-\Delta_\O}$ and $g(q,\bar q)$ is the regularized correlator on the pillow defined as
\begin{align}
g(q,\bar q) &\equiv \vev{  \O(0)\O(\pi)\O(\pi(\tau+1))\O(\pi\tau) } _{\mathbb{P}^1}\nn \\
&= \vev{ \psi | q^{L_0-c/24} \bar{q}^{\bar{L}_0-c/24}| \psi  }, \label{g-def}
\end{align}
with $\ket{\psi}= \ket{\O(\pi)\O(0)}_{\mathbb{P}^1}$. Equation \eqref{relate} together with \eqref{cross1} and the fact that $\vartheta_3(\tilde{q}) = \sqrt{-i \tau} \, \vartheta_3(q)$ imply that crossing symmetry is now a modular property for the pillow correlator \cite{MSZ}, 
\begin{align}\label{cross2}
g(\tau,\bar{\tau}) = (\tau \bar{\tau})^{\frac{c}{4}-4\Delta_\O} g(-1/\tau,-1/\bar\tau). 
\end{align}
This is the characteristic of a non-holomorphic modular form of weight $w=c/4-4\Delta_\O$.
 Additionally, it can also be seen from \eqref{g-def} that $g(q,\bar{q})$ decomposes into modified Virasoro blocks, $g(q,\bar{q}) = \sum_{h,\bar{h}} f^2_{\O \O \O_{h,\bar{h}}} \tilde{\cV}_h(q) \bar{\tilde{\cV}}_{\bar{h}}(\bar{q})$. The modified blocks, ${\tilde{V}_h}$ admit a $q$-expansion, 
\begin{align}
\tilde\cV_h(q)= \Lambda^{-1}(z) \cV_h(z) = (16)^{h-\frac{c}{24}} q^{h-\frac{c-1}{24}}  \eta(q)^{-{1\over 2}}H(h,q). \nn 
\end{align} 
The functions $H(h,q)$ are determined using the Zamolodchikov recursion relations \cite{Zamu}.
The appearence of the Dedekind-eta ($\eta(q)$) above suggests defining a normalized pillow correlator, which has a simpler $q$-expansion
\begin{align}  
p(q,&\bar{q}) =  g(q,\bar{q})\sqrt{ \eta(q) \eta(\bar{q})} = \sum_{h,\bar{h}} p_{h,\bar{h}}(q,\bar{q})\label{expansion1}  \\
&=\sum_{h,\bar{h}}  f^2_{\O \O \O_{h,\bar{h}}} 16^{\Delta- \frac{c}{12}}q^{h-\frac{c-1}{24} } {\bar{q}}^{\bar{h}-\frac{c-1}{24} } H(h,q) \bar{H}(\bar{h},\bar{q}). \nn 
\end{align}
Here, $\Delta = h+\bar{h}$ is the conformal dimension of the primary $\O_{h,\bar{h}}$. Using the modular property of $\eta(q)$ and \eqref{cross2}, we have the following modular crossing rule for our normalized pillow correlator,
\begin{align}\label{cross3}
p(q, \bar q)= \left(\tau\bar \tau\right)^{\frac{c-1}{4}-4\Delta_{\mathcal{O}}}p(\tilde q, \bar{\tilde{q}}).
\end{align}
This \red{non-holomorphic modular form relates the low energy CFT data to high energy asymptotics}. This shall be the central object for modular bootstrap in what follows. 

\begin{center}
	\begin{figure}[t]
		\label{pillow-msz}
		\includegraphics[scale=.3]{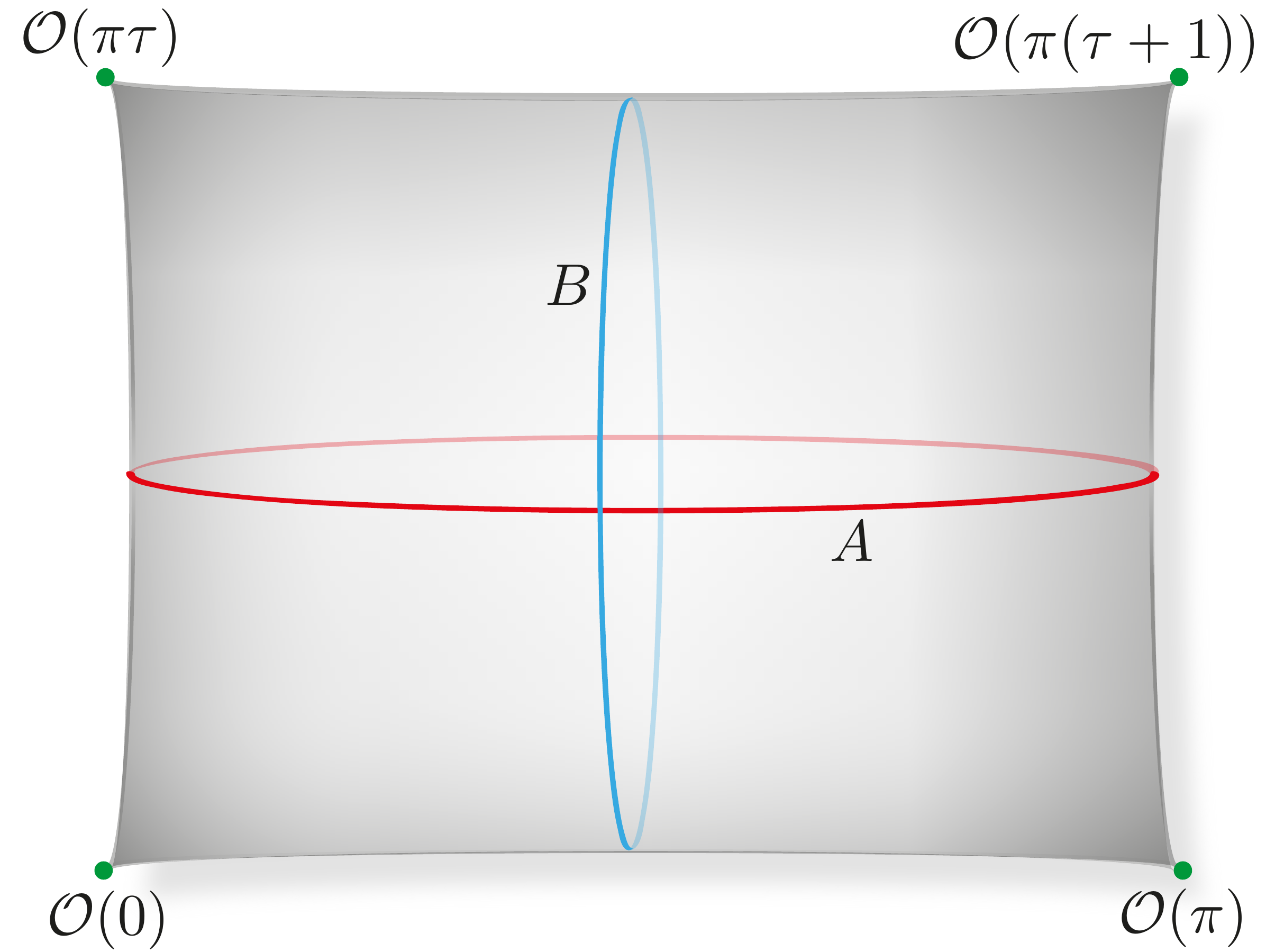}
		\caption{The pillow geometry $\mathbb{P}^1$ as a quotient of torus ($\mathbb{T}^2/\mathbb{Z}_2$). The CFT is quantized on the A-cycle (length $2\pi$) and propagation is along the B-cycle (which is halved compared to that of the original torus). The 4 dots are the fixed points of the orbifold at which the primaries are located. Crossing symmetry implies exchanging the A and B-cycles.}
	\end{figure}
\end{center}

\section{Extracting asymptotics \\ of OPE coefficients}
\label{sec:ope}
We will now use the $q$-expansion \eqref{expansion1} and the modular crossing property \eqref{cross3} of $p(q,\bar{q})$ to extract the OPE coefficient in the asymptotic limit of large intermediate conformal dimension $\Delta(=h+\bar{h})$ with dimension of external operators, $\Delta_\O$ fixed. The bootstrap equation \eqref{cross3} is true for all values of the modular parameter, $\tau$, lying on the upper-half plane. We shall now choose $\tau= i\beta/(2\pi)$, with $\beta \in \mathbb{R}^+$, without any loss of generality. $\beta$ therefore has a notion of an effective temperature parametrizing the pillow.

At low {temperatures}, i.e, $q = e^{-\beta/2}\rightarrow0$, the expansion  \eqref{expansion1} is dominated by the vacuum channel $h=\bar{h} = 0$. Furthermore, we have $H(q\to 0,h)\simeq 1+{\rm O}(q)$, which leads to $p(\beta\to \infty) \simeq 16^{- \frac{c}{12}}e^{\beta \frac{c-1}{24} }$. The modular property \eqref{cross3}  also implies the following equality,   $p(\beta) = \{\beta/(2\pi)\}^{\frac{c-1}{2} - 8\Delta_\O} p(4\pi^2/\beta)$, relating the high- and low-temperature expansions. Using the low temperature expansion for $p(4\pi^2/\beta)$, we thus obtain $p(\beta)$ at high temperatures,
\begin{align}\label{highT}
p(\beta\to 0) \  \simeq  \ 16^{- \frac{c}{12}}\left(\frac{\beta}{2\pi}\right)^{\frac{c-1}{2} - 8\Delta_\O}e^{\frac{\pi^2}{\beta} \frac{c-1}{6} }.
\end{align}

We now define the  \textit{weighted spectral density}, $\mathcal{K}_{\O}(\Delta) = \sum_{\alpha}f^2_{\O\O\O_{h_\alpha,\bar{h}_\alpha}} \delta(\Delta-\Delta_{\alpha})$. Here the sum is only over the primaries of the CFT. The sum in equation \eqref{expansion1} can then be rewritten as an integral
\begin{align} \label{int-rep-p}
\hspace{-.5cm}\int_{0}^\infty d\Delta ~ \mathcal{K}_{\O}(\Delta) 16^{\Delta-\frac{c}{12} } e^{-\frac{\beta}{2} \left( \Delta - \frac{c-1}{12} \right)} H(h,q)\bar{H}(\bar{h},\bar{q}).
\end{align}
Another drastic simplification arises from the heavy limit of the functions $H(h,q)$. In the heavy limit\footnote{{The heavy limit can derived from a monodromy analysis at large $c$ \cite{Kusuki1} such that $h | \log q|^2 \gg c$. On the other hand, \eqref{saddle} implies that $h | \beta_s|^2 \sim c$, thus for our analysis to be valid, we need an extended regime of validity of the heavy limit and this has been shown to be true indeed in \cite{Kusuki2}.} or in cases when the conformal blocks are known in closed form \cite{Nemkov}.}, we have $H(h\to \infty,q)\simeq 1+{\rm O}(h^{-1})$. The large $\Delta$ asymptotics of $\mathcal{K}_{\O}$ can now be expressed as an inverse Laplace transform of \eqref{highT} {:\begin{align}\label{inverse}
\mathcal{K}_{\O}(\Delta &\to \infty) \simeq \frac{1}{4\pi i (16)^{\Delta}} \int_{\epsilon-i\infty}^{\epsilon+i\infty} d\beta~  \left(\frac{\beta}{2\pi}\right)^{\frac{c-1}{2} - 8\Delta_\O} \nn  \\ 
&\times \exp\left(\frac{\pi^2}{\beta} \frac{c-1}{6}  +\frac{\beta}{2} \left( \Delta - \frac{c-1}{12} \right) \right).   
\end{align}
The above integral can be evaluated exactly, by utilizing the integral representation of the modified Bessel functions of the {first} kind,
\begin{align}
I_{\nu}(z)=\frac{1}{2\pi i }\left(\frac{z}{2}\right)^{\nu} \int_{\epsilon-i\infty}^{\epsilon+i\infty}\ dt\ \frac{1}{t^{\nu+1}}\exp\left[t+\frac{z^2}{4t}\right]. \nn 
\end{align}
where we identify, $\nu = 4\Delta_{\O} - \frac{c-1}{4}-\frac{1}{2}$ and $z = 2\pi \sqrt{\frac{c-1}{12} \left( \Delta - \frac{c-1}{12 } \right)}$. Thereby, we obtain}
\begin{align}\label{ans1}
\mathcal{K}_{\O}(\Delta\to\infty) &\simeq\frac{\pi}{16^\Delta} \left( \frac{ \Delta  }{\c }-1\right)^{ \nu } I_{2\nu }\left( 2 \pi \sqrt{ \c(\Delta-\c)}\right),
\end{align}
where $\c = (c-1)/12$ and $\nu = 4 \Delta_\O - 3\c-1/2$. 
One could have also proceeded to evaluate the inverse Laplace transform by finding the saddle of the integrand which is given by,
\begin{align}\label{saddle}
{\beta_s =   {\frac{2\pi }{\sqrt{\Delta/\c -1}}} + \frac{2 \nu +1}{\Delta - \c} + 
{\rm O}(\Delta^{-3/2}).}
\end{align}
This is consistent with the high temperature expansion \eqref{highT}, i.e.~the $\beta$ saddle is peaked around zero for $\Delta/\mathcal{C}\gg 1$. It can be very easily checked that the asymptotic expansion of \eqref{ans1} reproduces the result of the Laplace transform obtained via this saddle. The overall factor $(16)^{-\Delta}$ in \eqref{ans1} is unimportant 
as it gets eventually cancelled when multiplied with the Virasoro conformal blocks, see equation \eqref{expansion1} or \eqref{int-rep-p}. The conformal block also contains a factor $q^{\Delta}$, whose suppression is stronger than the growth of weighted spectral density \eqref{ans1} for heavy enough $\Delta$. This implies that the OPE converges \cite{RychkovOPE}. 

 The next step in the analysis is to note that the definition of $\mathcal{K}_{\O}(\Delta)$ can be used to naturally obtain an averaged three point coefficient squared, 
\begin{align}\label{avg}
\mathcal{K}_{\O}(\Delta) =  \sum\limits_{\alpha}f^2_{\O\O\O_{h_\alpha,\bar{h}_\alpha}} \delta(\Delta-\Delta_{\alpha}) = \rho(\Delta) \  \overline{f^2_{ \O \O\Delta}}
\end{align}
where, $\rho(\Delta) = \sum_\alpha \delta(\Delta-\Delta_\alpha)$, is the (unweighted) spectral density.
One can therefore estimate the average of the OPE coefficient as the ratio between $\mathcal{K}_{\O}(\Delta)$ and $\rho(\Delta)$. The large $\Delta$ asymptotics of $\rho(\Delta)$ is given by the Cardy formula \cite{Cardy:1986ie,Jabbari}\footnote{Note that, this is the density of states of primaries alone.}
\begin{align}\label{cardy}
\rho(\Delta \to \infty) &\simeq 2\pi I_{0}\left( 4\pi \sqrt{ \c (\Delta - \c)}\right).
\end{align}
As is well known, this expression is obtained by evaluating an inverse Laplace transform. If the integral is done by the saddle-point method and the leading fluctuations are included, then one can once again check agreement with the expansion of the Bessel function appearing in \eqref{cardy},
\begin{align}\label{cardy-ex}
 \rho(\Delta\to\infty) &\simeq \frac{1}{\sqrt{2\c}}\left( \frac{\Delta}{\c}-1 \right)^{-1/4} e^{ 4 \pi \sqrt{ \c ( \Delta -\c )}}. 
 \end{align}
The  statistical entropy, $S(\Delta)$, can be  obtained as usual from the logarithm of the density of states. 
\begin{align}\label{entropy-01}
S(\Delta \to \infty) &\simeq S^{(0)} (\Delta) + S^{(1)} (\Delta)  +\cdots \\ &= { 4 \pi \sqrt{ \c ( \Delta -\c )}} - \frac{1}{4} \log \left[ \c (\Delta -\c)  \right] +\cdots.  \nn 
\end{align}


Finally we arrive at the asymptotic expression for the mean-squared OPE coefficient, from \eqref{avg} using \eqref{ans1} and \eqref{cardy}
\begin{eqnarray}\label{answer}
&{}&\overline{f^2_{ \O \O\Delta}} = \frac{\mathcal{K}_{\O}(\Delta)}{\rho(\Delta)}  \\  &\underrel{\Delta \rightarrow \infty}{\simeq}& \frac{16^{-\Delta}}{2} \left( \frac{ \Delta}{\c }-1\right)^{ \nu}\frac{ I_{2\nu }\left( 2 \pi \sqrt{ \c(\Delta-\c)}\right)}{I_{0}\left( 4\pi \sqrt{ \c (\Delta - \c)}\right)} . \nn 
\end{eqnarray}
The leading dependence on $\Delta$ can be obtained from the asymptotic form of the modified Bessel functions. This gives
\begin{align}\label{answer-ex}
\overline{f^2_{\O\O\Delta}} &\underrel{\Delta \rightarrow \infty}{\approx}& \frac{16^{-\Delta}}{\sqrt{2}} \left( { \Delta /\c -1  }\right)^{\nu } \exp \left[{-{S^{(0)}(\Delta) \over 2}}\right], 
\end{align}
where $S^{(0)}(\Delta)={ 4 \pi \sqrt{ \c ( \Delta -\c )}}$ is the leading term in the entropy  \eqref{entropy-01}.
The above result shows that the \textit{mean} squared coefficient  has an entropic suppression\footnote{The factor of $16^{-\Delta}$ does not survive once the conformal block is multiplied.}. {A potentially more rigorous derivation of the above result can be performed by using Tauberian theorems which would relate $\mathcal{K}_{\mathcal{O}}(\Delta)$ and $p(q,\bar{q})$  \cite{Mukhametzhanov:2018zja,Qiao:2017xif}.} \red{This behaviour has been predicted earlier for CFTs in arbitrary dimensions using phase space arguments and holographic expectations in \cite{RychkovOPE,FK-S-matrix}}. We shall explore further consequences of this suppression in the next section. 
%

\section{The gravitational S-matrix and black holes}
\label{sec:bh}

\red{We shall now specialize to holographic CFTs with large central charge and having a sufficiently sparse spectrum of light operators to allow a Hawking-Page transition between thermal $AdS_3$ and BTZ geometries \cite{Heemskerk:2009pn,ElShowk:2011ag,Hartman:2014oaa,Haehl:2014yla,Belin:2014fna}.} We additionally require the light density of states to be sparse and assume that the validity of the analysis of the previous section can be extended to the regime where $c\to \infty$ with $\Delta/c \sim {\rm O}(1)$ analogous to \cite{Hartman:2014oaa} for applicability to the black hole regime.
The average over heavy primaries leads to the notion of a black hole in the holographic dual \cite{Carlip:2000nv,km}. \red{Holography, therefore, provides a natural framework for applying the analysis of the previous  section. }

Let us return to the integral representation of the function $p(q,\bar{q})$ in \eqref{int-rep-p}.  It is expected that the small $\beta$ behavior (or the high temperature regime equivalent to $q\to1$ or $z\to 1$) will be governed by a saddle point of high conformal dimension.\footnote{\nuen{Equivalently, if one considers the t-channel, the dominant contribution is from the vacuum and light states. The two descriptions are related by modular transformations  of the pillow.}} Also due to the exponential growth in the density of states with energy \eqref{cardy-ex}, in this regime the full 4-point correlator, $\cF(z,\bz)$ is dominated by partial wave amplitudes corresponding to heavy intermediary channels, $\cF_{\Delta}(z,\bz)$. {Owing to the coarse-graining over heavy states, this corresponds to a classical black hole in the intermediate state of a $2\to 2$ scattering process in the bulk.} We shall use our formula for the mean-squared OPE coefficient \eqref{answer}, to arrive at a typical estimation of $\cF_{\Delta}(z\to 1,\bz\to 1)$, {where $\Delta$ represents the state averaged over all heavy primaries}. 

The averaged {contribution} from  heavy pillow blocks, in the decomposition \eqref{expansion1}, is (we can once again set the recursion factors, $H(h,q)$ and $\bar{H}(\bar{h}, \bar{q})$ to 1) 
\begin{align} \label{ppredx}
p_{\Delta\to \infty}(q,\bar{q}\to 1) &\simeq \frac{16^{-\c}}{{2}^{5/6}}\left( { \Delta/\c - 1 }\right)^\nu e ^{{-{S^{(0)}(\Delta) \over 2}}} .
\end{align}
Finally the leading behaviours, in the limit $z\to 1$ or $\tq \to 0$, of $\Lambda(z)$ and Dedekind-eta appearing in $\cF_\Delta(z,\bz)$ are dictated by their S-modular transformation properties. 
Combining  these ingredients  together gives the dominant contribution (mediated by heavy exchanges) to the 4-point function in the limit $z\to 1$,  (using the notation of \eqref{cF})
\begin{align}\label{finite-c-4}
\cF_{\Delta}(z\to 1,&\bz \to 1) \simeq \, \left| \frac{1}{\pi} \log \frac{16}{1-z} \right|^{6\c - 8\Delta_\O} | 1-z |^{\c - 2\Delta_\O} \nonumber \\
&\times \frac{ \left( { \Delta /\c -1  }\right)^{ 4 \Delta_\O - 3\c-1/2 }}{2^{4\c + \tfrac{1}{2}}} \exp \left[-2\pi \c \sqrt{\Delta - \c} \right]. 
\end{align}
%
\red{At large central charge}, the suppression for high values of $\Delta$ in the above formula is precisely captured in terms of the Bekenstein-Hawking (BH) entropy, $S_{\rm BH}$, of the BTZ black hole (bh). Defining this suppression factor as $\mathcal{Y}(\Delta)$, \red{ using $\c\simeq c/12$ and ignoring logarithmic corrections in $\Delta$, we have}
\begin{align}
\mathcal{Y}(\Delta) = \exp\left[-2\pi\sqrt{{c\over 12}\left(\Delta -{c\over 12 }\right)}\, \right]  \ = \ \exp \left[- \frac{S_{\rm BH}}{2}  \right]. 
\end{align}
This forms the key result of this paper. 
Quite remarkably, it has been long expected that this heavy-regime should be dominated by black hole exchange in the gravity dual. In fact, it has been shown that high-energy $2\to 2$ scattering processes (with small impact parameter) containing black holes as intermediate states have the following $\mathcal{S}$-matrix \cite{GiddingsPorto,FK-S-matrix}\footnote{Strictly speaking one would need to take the flat spacetime limit to obtain a notion of in and out states leading to definition of a S-matrix. This affects the kinematic/cross-ratio dependence but not the OPE coefficient.}
\begin{align}\label{holy-grail}
\mathcal{S} \sim \exp \left[- \frac{S_{\rm BH}}{2}  \right].
\end{align}
The arguments for this behaviour are on fairly general grounds based on black hole thermodynamics. The size of the phase space of black hole microstates is given by $e^{S_{\rm BH}}$. Therefore, the cross-section or the expected probability for production of a black hole microstate is $e^{-S_{\rm BH}}$. {The probability of the inverse process, corresponding to the black hole evaporating into two scalars is given by the same factor $e^{-S_{\rm BH}}$, due to time reversal invariance. Finally, in order to describe the black hole classically, we multiply by all possible black hole states $e^{S_{\rm BH}}$.\footnote{We are very grateful to Per Kraus for explaining this to us.} This is}
\begin{align}
{|\mathcal{S}|^2 \sim \underset{\text{bh forms}}{e^{-S_{\rm BH}}} \ \times  \  \underset{\text{bh decays}}{e^{-S_{\rm BH}}} \ \times  \  \underset{\text{degeneracies}}{e^{S_{\rm BH}}} , }
\end{align}
{which leads to \eqref{holy-grail}.}
As mentioned earlier, conformal field theory furnishes a microscopic description of a  black hole by  coarse-graining over a family of heavy primaries. Our analysis therefore provides a microscopic derivation of this feature \eqref{holy-grail} and, at the same time, goes beyond the semi-classical approximation (eq.~\eqref{finite-c-4} is non-perturbative in the Newton's constant $G_N={3\over 2c}$). 

\red{We should emphasize, however, that  the exponentially decaying feature of the averaged heavy channel is more generally true for all CFTs having Virasoro symmetry with $c>1$, \ie regardless of AdS/CFT or any specific restrictions on the spectrum. Moreover, this aspect  implies convergence of the conformal block expansion \cite{RychkovOPE,FK-S-matrix}. }
 




\section{Conclusions}
\label{sec:conclusions}

In this work we have derived an universal formula for the mean-squared 3-point coefficient, $\overline{f_{\O\O \Delta}^2}$, wherein the averaging for one of the operators is performed over heavy primaries. This information was then used to determine the contribution of heavy primaries to the 4-point function of identical scalars. We found that these contributions are entropically suppressed. \red{Interpreted holographically,} this implies a behaviour ${\mathcal{A} } \sim e^{-\pi \sqrt{ \frac{\Delta  c}{3} }} \sim e^{-S_{BH}/2}$ for the gravitational $2\to 2$ $\cS$-matrix which leads to black hole formation and evaporation. Our analysis essentially utilized the mapping of the sphere to the pillow geometry and modular properties of the correlator therein \cite{MSZ}. 

In terms of Mandelstam energy variable $s$, the scattering amplitude behaves like ${\mathcal{A} } \sim \exp(-s^{1/4})$. In the context of locality of quantum field theories our result is well over the Cerulus-Martin lower bound ${\mathcal{A} } \geq e^{-\sqrt{s } \log s }$ \cite{Cerulus} implying that the involved interactions are local. It may also be interesting to think of this amplitude arising from string scattering. It is known that at the tree level the amplitudes behave extremely softly $\sim e^{-s \log s}$ \cite{Veni}\footnote{The violation of the Cerulus-Martin bound exhibits the non-local nature of the string \cite{Gross87}.}, however higher order corrections change this behaviour to $e^{-\sqrt{s}}$ restoring the Cerulus-Martin bound \cite{Ooguri}. This result comes from a Borel resummation of string perturbation theory and is expected to be true in a high energy window, $ \log(1/g^2) < s < \left(\log(1/g^2)\right)^3$, where $g$ is the string coupling. It will therefore be interesting to obtain the increased amplitude, $\mathcal{A}$, from the regime $s > \left(\log(1/g^2)\right)^3$.

\red{Since no systematic classification exists for the space of CFTs with $c>1$, it is worthwhile to extract the similarities and differences between these theories. The work \cite{Cardy:1986ie} and \cite{km} along with the present one uncovers the features of the spectra and OPE coefficients which these theories have in common.}
We hope that this work advances the program of the conformal bootstrap for these theories by opening up the avenue of bootstrapping via mapping to the pillow geometry. It would be interesting to realize the potential of this mapping further, with a view towards extracting statistics of OPE coefficients in the light and intermediate regimes (in the spirit of \cite{Hartman:2014oaa,Kraus:2017kyl}). 
\red{The results can be straightforwardly generalized for off-diagonal mean-squared OPE coefficients of the kind $\overline{f^2_{\mathcal{O}_1 \O_2 \Delta}}$.}
 Additional simplifications may also arise in the semi-classical (large $c$) regime in which the conformal block exponentiates \cite{Fitzpatrick:2014vua,Chang:2015qfa,Kim:2015oca,Chang:2016ftb}. 

Finally, it is intriguing to note that mean-squared statistics of all the three operators being heavy in $f^2_{abc}$ has been derived using the genus-2 modular bootstrap \cite{CardyMaloney}. In a sense, this is consistent with our result as each heavy index, $a$, contributes via a factor of $e^{-S(\Delta_a)/2}$.  
It would be worthwhile to formulate a more direct approach to get finer statistics of the OPE coefficients from higher genus \cite{CardyMaloney,Keller:2017iql,Cho:2017fzo}.

\section*{Acknowledgments}
 It is a pleasure to thank Sumit Das, Justin David,  Gia Dvali,  Matthias Gaberdiel, Tom Hartman, Jared Kaplan, Ivo Sachs and especially Per Kraus for fruitful discussions.
 DD acknowledges the support provided by the Alexander von Humboldt Foundation and the Federal Ministry for Education and Research through the Sofja Kovalevskaja Award. DD also thanks ETH Z\"urich for hospitality where a part of this work was completed. The work of SD is supported by the NCCR SwissMAP, funded by the Swiss National Science Foundation. In addition, SD thanks   JMU W\"urzburg,  LMU M\"unchen and IISc Bangalore for hospitality during various stages of the work. SP acknowledges
the support provided by the US Department of Energy (DOE) under cooperative research
agreement DE-SC0009919.
%


\appendix*
\section{Elliptic representation}
\label{supp-1}
The pillow frame $\mathbb{P}^1\equiv\mathbb{T}^2/\mathbb{Z}_2$ is constructed by orbifolding a torus. Using the knowledge of the marked points of the sphere in $x$,  we can define a coordinate $u$ on the torus,
\begin{align}
du = \frac{1}{\vartheta_3(q)^2} \frac{ dx}{ \sqrt{ x(z-x)(1-x)}}.
\end{align}
The integrals along the branch cuts give the cycle lengths  and are the origin of the $K(z), K(1-z)$ functions in the definition of the modular parameter. In the pillow frame, the CFT has a natural interpretation of being quantized along one cycle, while evolving along the other (see Fig.~1). The factor $\vartheta_3(q)^{-2}$ normalizes our quantization cycle (the A cycle) to $2\pi$. The $\mathbb{Z}_2$ symmetry acts as $u\rightarrow-u$, so the propagation on the B-cycle is halved. This is reflected in the nome $q = e^{ i \pi \tau}$, as opposed to $q = e^{2 \pi i \tau}$ for the unorbifolded torus. The four marked points on the sphere get mapped to the four fixed points of the orbifold.

In Section IV we are interested in the $z\rightarrow 1$ limit of the $s$-channel. The modular parameter of the pillow in this regime is 
\begin{align}\label{limittau}
\tau(z\to 1) \simeq i \pi \left(  \log \frac{16}{1-z} \right)^{-1}.
\end{align}
This implies, that $\tilde{q} = e^{-\frac{\pi i }{\tau} } \rightarrow 0.$


\bibliographystyle{apsrev}
\bibliography{pillowref}

\end{document}